# A Low-Cost Machine Learning Based Network Intrusion Detection System with Data Privacy Preservation

Jyoti Fakirah [1], Lauhim Mahfuz Zishan [1], Roshni Mooruth [1], Michael N. Johnstone [2], and Wencheng Yang[2]

[1] School of Science, Edith Cowan University, Joondalup, Western Australia

[2] Security Research Institute, School of Science, Edith Cowan University, Joondalup, Western Australia
{m.johnstone, w.yang}@ecu.edu.au

**Abstract.** Network intrusion is a well-studied area of cyber security. Current machine learning-based network intrusion detection systems (NIDSs) monitor network data and the patterns within those data but at the cost of presenting significant issues in terms of privacy violations which may threaten end-user privacy. Therefore, to mitigate risk and preserve a balance between security and privacy, it is imperative to protect user privacy with respect to intrusion data. Moreover, cost is a driver of a machine learning-based NIDS because such systems are increasingly being deployed on resource-limited edge devices. To solve these issues, in this paper we propose a NIDS called PCC-LSM-NIDS that is composed of a Pearson Correlation Coefficient (PCC) based feature selection algorithm and a Least Square Method (LSM) based privacy-preserving algorithm to achieve low-cost intrusion detection while providing privacy preservation for sensitive data. The proposed PCC-LSM-NIDS is tested on the benchmark intrusion database UNSW-NB15, using five popular classifiers. The experimental results show that the proposed PCC-LSM-NIDS offers advantages in terms of less computational time, while offering an appropriate degree of privacy protection.

## 1 Introduction

Cybercriminals are often first users of new network technologies which enables their attempts to compromise networks to be successful, thus defenders are at a disadvantage. Machine Learning is potentially a way to redress this imbalance. A Network Intrusion Detection System (NIDS) is a device or program that scrutinises the (protected) systems for malicious behaviours and generates warnings when a cyber-attack takes place. A NIDS can perform several actions including examination and assessment of system and user actions, evaluation of system and data records , audit of (operating) system weaknesses and configurations and analysis of unusual/untrustworthy activity [1]. A NIDS is usually located at critical network points such as gateways or routers to gain access to network traffic. The NIDS supervises and pinpoints network-attack models over networking environments and safeguards computing resources against malevolent events. A NIDS can be characterised by the detection method it applies, for example, misuse detection and anomaly detection [2]. Anomaly-based systems are popular due to their ability to discover new types of threats. Unfortunately, these types of systems suffer from large numbers of false positive alerts, that is many normal packets can be misclassified as attack packets [3].

Machine Learning (ML) techniques, e.g., Random Forest (RF), Decision Tree (DT) and K-Nearest Neighbors (KNN), are applied in many domains including NIDS. As ML algorithms are data-driven, and NIDSs are no exception, they require significant amounts of training data to classify new data effectively (correctly). This inevitably causes some sensitive network data to be exposed, therefore, a privacy preservation approach is required to conceal sensitive data, whilst maintaining the effectiveness of the ML algorithm in the NIDS. To examine the weaknesses in privacy preservation, various studies have suggested privacy attack techniques such as minimal attacks, and contextual knowledge attacks. To overcome these attacks, $l$-diversity, $t$-closeness, $k$-anonymity, and the least square method are proposed. It is vital to safeguard data during the training stage of the ML techniques [4].

Moreover, in some applications, the ML-based NIDS operate on edge devices with limited resource, thus low cost (compute or power) is an essential requirement of the NIDS. Feature selection is a data reduction technique embraced by ML researchers to reduce the "curse of dimensionality" through the elimination of less important attributes. The selection of a subset of significant features from the entire dataset often results in enhanced model performance, accuracy and interpretability and reduced computational cost of resource-limited devices [5].

Motivated by the above issues, in this paper we propose a machine learning-based network intrusion detection system called PCC-LSM-NIDS that can achieve low-cost intrusion detection while providing privacy preservation of sensitive data. Specifically, a Pearson Correlation Coefficient (PCC) based feature selection algorithm is utilized to extract the optimal subset of key attributes from an original dataset to reduce the amount of data in the ML training and testing phases and save precious computational resource, thus the low-cost criterion is achieved. The application of a Least Squares Method (LSM) [6] based data privacy preservation algorithm to the extracted optimal attributes of the original dataset protects the original sensitive data by convert-ing/distorting. The converted/distorted data are then run into five widely used ML classifiers, specifically Random Forest (RF), Decision Trees (DT), Naïve Bayes (NB), Support Vector Machine (SVM) and K-Nearest Neighbors (KNN). The experiments are carried out on the UNSW-NB15 dataset [7].

The rest of this paper is constituted as follows: Section 2 portrays related work on IDSs, privacy-preserving techniques, feature selection and machine learning. The proposed PCC-LSM-NIDS is detailed in section 3. Section 4 describes the experimental results and presents a discussion. Finally, Section 5 concludes this research work.

## 2    Related Works

A study by Aravind et al. [8] suggested an Intrusion Detection System (IDS) based on a K means classifier, using the UNSW NB-15 dataset Their method claims 90% accuracy for the attacks. Kamarudin et al. [9] described an anomaly-based IDS that utilises an ensemble classification method to identify anonymous and known attacks on websites. The procedure requires eliminating unrelated and unnecessary attributes, making use of a filter and wrapper selection process to acquire the most important variable. A data mining method is employed using the boosting algorithm Logitboost with a Random Forest algorithm to attain high detection precision whilst maintaining a low false alarm rate. The IDS was assessed using the NSL-KDD and UNSW NB15 datasetsFor the NSL-KDD dataset, they reported a false alarm rate of 8.22%, a detection rate of 89.75% and accuracy of 90.33%. For the UNSW NB-15 dataset the results were as follows: a false alarm rate of 0.18%, a detection rate of 99.10% and an accuracy rate of 99.45%.

A framework was recommended by Beloucha et al. [10] which assesses the performance of four machine learning classifiers specifically: Decision Tree, Naïve Bayes, Random Forest and SVM by utilising Apache Spark for intrusion detection in network traffic. Apache Spark is a cluster computing platform and is intended to embrace a huge collection of tasks that needed individual distributed systems earlier. The UNSW NB-15 dataset was used. The job of the detection classifier was to categorize if the incoming traffic was normal or abnormal. It is noted that the Random Forest algorithm performed well compared to the other algorithms with respect to True Positives. The RF achieved a recall of 93.53% followed by Decision Tree with 92.52%. SVM and Naïve Bayes reported similar True Positives of 92.46% and 92.13% respectively. The researchers discovered that True Negative Rate for RF and DT based schemes were effectively identical at 97.75% and 97.10% respectively. The True Negatives for SVM was 91.15% and the Naïve Bayes classifier did not perform well in terms of specificity. RF performed well in terms of accuracy with 97.49% whereas Naïve Bayes has the lowest accuracy at 74.19%. Beloucha et al. concluded that RF yielded the best performance in terms of recall, true negative rate, and accuracy.

Zhou et al. [11] recommended Deep Feature Embedding Learning (DFEL) to detect intrusions on the Internet of Things (IoT). It is claimed that DFEL balances detection performance and speed. The UNSW-NB15 dataset is divided by following the identical procedure. To fit DFEL, 80% of the data was employed and the pre-trained prototype was obtained. The other 20% was divided into 70%/30% as training and testing for ML algorithms. Afterwards, by making use of the DFEL, the 20% data are transferred to latent features and the embedding variables are divided into 70%/30% for embedding training and testing. At last, the results from the machine learning classifications are evaluated on embedding data and initial data. Gradient-boosted trees (GBT), KNN, DT, NB, SVM and logistic regression (LR) are applied for boosting the detection speed. The DFEL method improves most algorithms accuracy and substantially conserve the cyber detection time. The performances of these classifiers are assessed with or without DFEL. The accuracy for the classifiers is as follows: NB 92.52%, KNN 91.90%, DT 92.29%, LR 92.35%, SVM 92.32% and GBT 93.13%. It can be noted that GBT has attained the highest accuracy compared to other algorithms and there is a rise in Precision and Sensitivity for the proposed classifiers with DFEL.

Mandala et al. [6] studied the Least Square Privacy Preservation Method distortion technique using Least Square Method with ensemble classification method for delivering enhanced privacy preservation on intrusion data. The accuracy before and after distortion was tested using the WEKA tool and Java code for measuring privacy parameters. While making a comparison with the performance of the techniques provided in WEKA (NB, SMO, IBK and J48), it was observed that they have obtained the same outcomes to original accuracy, False Alarm Rate and F-score. It was observed that the implementation of LSPPM-NIDS for intrusion detection diminished the whole computational time with minimal loss of information.

Keshk et al. [12] proposed a new Privacy Preservation Intrusion Detection (PPID) method using the Pearson Correlation Coefficient for choosing essential data without compromising sensitive information of SCADA data. This study made use of the EM clustering algorithm for detecting intrusive observations of SCADA instances. Attributes were chosen based on correlation coefficient opting for sections with less sensitive information of the SCADA data. Afterwards, the EM clustering algorithm assembled SCADA data to identify abnormal behaviours successfully. It was observed that reducing the number of variables avoided revealing sensitive information and to some extent decreased the detection rate of attacks.

## 3       The Proposed System

To achieve low cost network intrusion detection and provide privacy preservation of sensitive data, the proposed PCC-LSM-NIDS includes a PCC-LSM model, which is composed of two algorithms: 1. A PCC-based feature selection algorithm to select an appropriate subset of features results in a simpler modelling process with possible more accurate intrusion detection rates within a shorter span of time, giving the fact that high dimensional data usually consist of redundant and less important features often affects the performance accuracy of a detection model as well as computation time and cost. 2. The LSM-based data privacy preservation algorithm to convert/distort the selected dataset into another version in a non-invertible manner.

### 3.1     Pearson Correlation Coefficient (PCC) based feature selection

Given a dataset, e.g., UNSW-NB15 dataset, a Pearson Correlation Coefficient (PCC) PCC based feature selection algorithm is applied to select key attributes, based on the strength of linear dependency between the dependent and independent attributes. The primary objective for choosing a portion of the features, so as to compensate the time taken for applying data distortion techniques in the model and reduce the computational cost.

The PCC of the two features $f_1 = [x_1, x_2, \ldots x_N]$ and $f_2 = [y_1, y_2, \ldots y_N]$ is calculated [13] as,

$$PCC(f_1, f_2) = \frac{cov(f_1, f_2)}{\sigma_{f_1}\sigma_{f_2}} = \frac{\sum_{i=1}^{N}(x_i - M_{f_1})(y_i - M_{f_2})}{\sqrt{\sum_{i=1}^{N}(x_i - M_{f_1})^2}\sqrt{\sum_{i=1}^{N}(y_i - M_{f_2})^2}} \qquad (1)$$

where $cov(\ )$ is the covariance and $\sigma$ is the standard deviation; $M_{f_1}$ and $M_{f_2}$ are the mean of features $f_1$ and $f_2$.

For the ranking of the strongest attributes, the mean of each PCC feature $M_{pcc_{f_i}}$ is calculated as,

$$M_{pcc_{f_i}} = \frac{1}{N}\sum_{i=1}^{N}PCC_{f_i} \qquad (2)$$

Subsequently, the means are arranged in descending order to establish strongly related features.

### 3.2    Least Square Method (LSM) based data privacy preservation

To preserve sensitive data, the original data in the dataset are distorted using the Least Squares Method (LSM) [6]. LSM is an algebraic approach based on the assumption that there exists a linear relationship between the target variable and independent variables which are expressed as follows:

$$Y_n = \beta_0 + \beta_1 X_1 + ... + \beta_n X_n + \varepsilon \tag{3}$$

where $X$ is independent variables, $Y$ is the target variable, $\beta$ (beta) is parameter estimates (coefficients) and $\varepsilon$ (epsilon) is the residual error.

Given a dataset $D$ of $n$-dimensions is changed into numeric, where the nominal features are substituted with integer values. The numeric dataset is viewed as matrix $X$ in the size of $n*m$, which is successively transformed as matrix $TX$. The values in the original matrix $X_{n*m}$ are altered using the LSM to obtain $TX_{n*m}$ matrix where $n$ and $m$ designate the number of rows and columns in the matrix.

For the linear equation,

$$X\beta = Y \tag{4}$$

The solution vector is then computed as follows:

$$\beta = X^{-1}Y \tag{5}$$

Given the different sizes of $X$ and $Y$, Equation (5) is adjusted to Equation (6) to solve for the solution vector.

$$\beta = (X^T X)^{-1} X^T Y \tag{6}$$

where, $\beta = [\beta_0, \beta_1, \beta_2, ..., \beta_n]$ is the solution vector for the original matrix $X$ and $\beta_0$ is the intercept value ($c$) for the prediction of the training dataset. The residual error $\varepsilon$ is calculated as illustrated in the equation below,

$$\varepsilon = \tfrac{1}{n}\sum\nolimits_{i=1}^{n}(Y_i - y_i)^2 \tag{7}$$

where, $Y_i$ is the actual value and $y_i$ is the predicted value.

The transformed dataset *TX* is then achieved by multiplying the coefficients of the solution vector ($\beta$) and each column value in the matrix X and add the intercept and residual error values, in the same way that transfers the element $X_{11}$ to $TX_{11}$ as shown in Equation (8),

$$TX_{11} = (\beta_1 * X_{11}) + (c + \varepsilon) \tag{8}$$

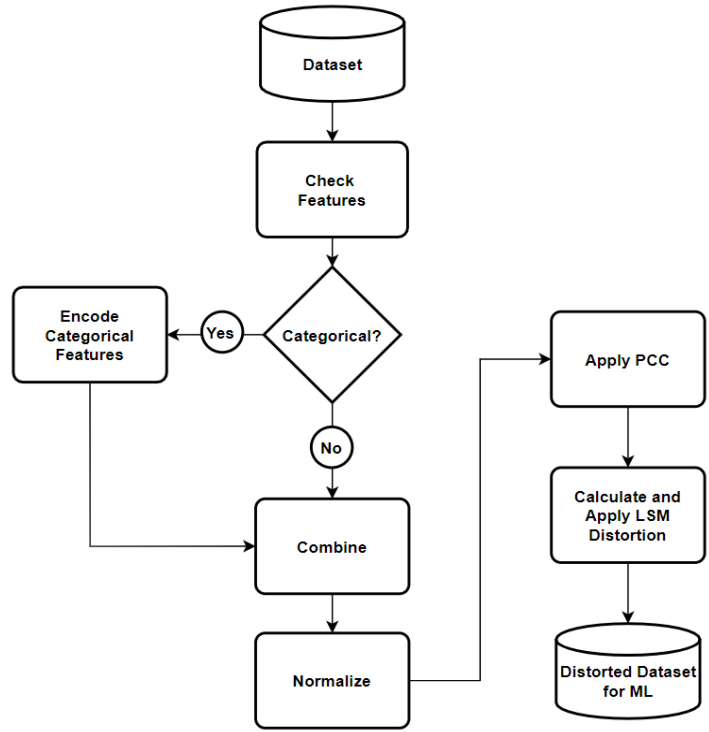

**Fig. 1.** Data selection and distortion process of the proposed model.

Fig.1 details the overall data selection and distortion process of the proposed PCC-LSM model. The transformed data is then fed as input to the ML classifiers to detect intrusions in network traffics. In this work, five popular classifiers, namely Support Vector Machine (SVM), Naïve Bayes (NB), Decision Tree (DT), Random Forest (RF) and K-Nearest Neighbors (KNN), are employed and the performance of the proposed PCC-LSM-NIDS is evaluated using Recall, Precision, Specificity, F-Score and Accuracy metrics.

## 4 Experimental Results and Discussion

All the experiments were performed in a test environment with an AMD Ryzen 3600x 6 cores 12 threaded processor, clocked at around 4.4GHz, 16GB of 3600MHz DDR4 RAM and 1TB of NVME SSD storage running Python 3.8 on Spyder IDE. The operating system is Windows 10 Pro, OS Build 19041.572. The average runtime was computed without any background task running and an average of 3-4 readings was taken for each instance. So, avg. runtime specified in the results was considered as a standard about a similar hardware and software setup.

### 4.1 Dataset Selection

UNSW_NB15 dataset is the latest publicly available dataset, initiated in 2015 by the Cyber Range Lab of the Australian Centre for Cyber Security (ACCS) for exploration of intrusion detection systems [14]. It was created using four tools namely AXIA Perfect Storm, Tcpdump, Argus and Bro-IDS, to generate a fusion of contemporary usual network activities and synthetic modern attack behaviours from network traffic. The dataset contains 49 features across 175,341 training and 82,332 testing data records correspondingly. The partitioned datasets consist of 45 attributes, where features 'srcip', 'sport', 'dstip' and 'dsport' are excluded. Given the difference in levels for categorical variables in the training and testing datasets, only the training dataset CSV file containing 175,341 records is used to investigate the proposed PCC-LSM-NIDS to reduce bias in the performance accuracy [14]. The first attribute relates to the id which is removed before any analysis. Each data record has 44 features categorised into Flow, Basic, Content, Time, Additional General and Label features shown in Table 1. Feature 48 represents the label of each record while feature 49 represents the attack categories. There are 9 attack categories: Fuzzers, Analysis, Backdoors, Denial of Service, Exploits, Generic, Reconnaissance, Shellcode and Worms.

**Table 1.** Summary of UNSW-NB15 feature categories.

| Features | Description |
| --- | --- |
| Flow (5) | proto |
| Basic (6-18) | state, dur, sbytes, dbytes, sttl, dttl, sloss, dloss, service, sload, dload, spkts, dpkts |
| Content (19-26) | swin, dwin, stcpb, dtcpb, smeansz, dmeansz, trans-depth, res_bdy_len |
| Time (27-36) | sjit, djit, stime, ltime, sintpkt, dintpkt, tcprtt, synack, ackdat, is_sm_ips_ports |
| Additional General (37-47) | ct_state_ttl, ct_flw_http_method, isftp_login, ct_ftp_cmd, ct_srv_src, ct_srv_dst, ct_dst_ltm, ct_src_ltm, ct_src_dport_ltm, ct_dst_sport_ltm, ct_dst_src_ltm |
| Label (48-49) | label (0 for normal and 1 for attack records) |

| | attack cat (Nine attack categories: Fuzzers, Analysis, Backdoors, Denial of Service, Exploits, Generic, Reconnaissance, Shellcode and Worms) |
|---|---|

## 4.2    PCC Analysis

In the experiment, a correlation coefficient threshold of 0.85 is applied where features with correlation strength above 0.85 are dropped before any further analysis. The PCC technique dropped 17 attributes from each data record in the UNSW-NB15 network intrusion dataset. The dropped features are, 'ct_srv_dst', 'synack', 'ct_src_dport_ltm', 'is_sm_ips_ports', 'dwin', 'sloss', 'ct_dst_src_ltm', 'sbytes', 'ct_src_ltm', 'ct_dst_sport_ltm', 'dloss', 'dbytes', 'ack-dat', and 'ct_ftp_cmd' along with three categorical features, 'proto', 'state' and 'service'. Fig. 2 illustrates the graphical representation of the PCC correlation matrix of some features with coefficient values denoted by the color scale.

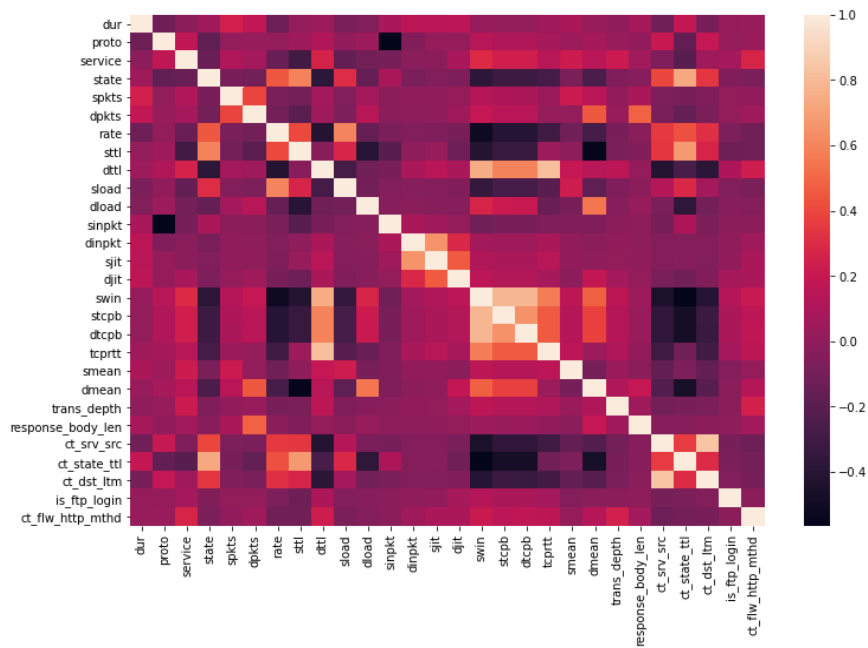

**Fig. 2.** Heatmap of feature correlations with PCC.

### 4.3    Performance Evaluation

The performance of the proposed PCC-LSM-NIDS is gauged using the conventional True Positive (TP), False Negative (FN), False Positive (FP) and True Negative (TN) measures [15]. The performance of the proposed system using different classifiers, RF, DT, NB, SVM and KNN, are compared.

Fig. 3 represents the comparative analysis of classifiers' performance. From the experimental results in Fig. 3, it can be seen that the RF, DT and KNN classifiers perform better than the NB and SVM classifiers with the proposed PCC-LSM model in terms of Specificity, Precision, Accuracy and F-Score values, except Recall.

Fig. 4 and Fig. 5 show the classifiers' computation time (training and testing time) before and after implementing the PCC-LSM model, respectively. A major drop-in computation time can be observed with the proposed PCC-LSM-NIDS. Hence, it can be concluded that the proposed model is economical in terms of computation time and cost.

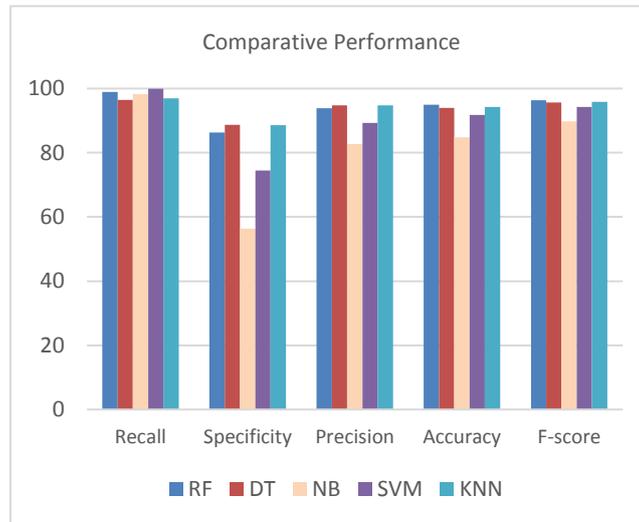

**Fig. 3.** Comparative classifier performance.

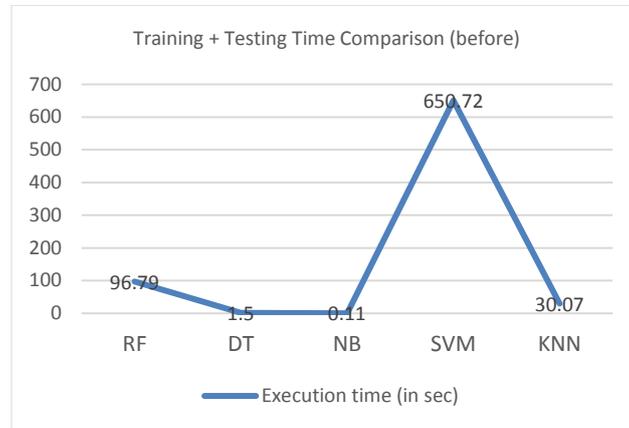

**Fig. 4.** Classifier computation time (before).

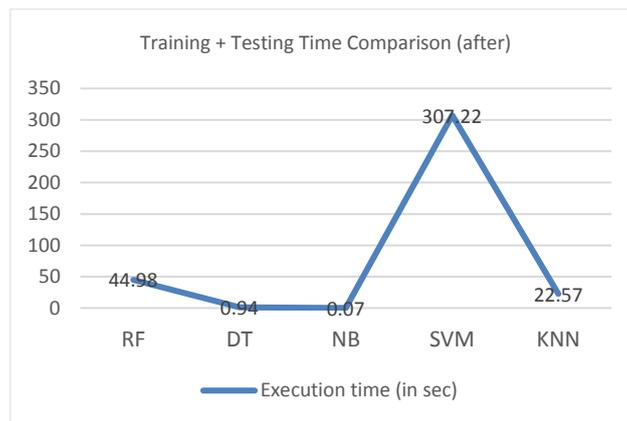

**Fig. 5.** Classifier computation time (after).

### 4.4 Privacy Analysis

Data Utility (DU) and Privacy Measures (PMs) are two important metrics to evaluate the efficiency of the proposed PCC-LSM-NIDS. The assessment of classifiers' performance is done before and after the implementation of PCC and LSM techniques. Usually, a data conversion influences the behaviour of the original data and so, DU and PMs are computed to measure this behavioural change. DU measures the accuracy between the original and transformed datasets.

The relative value difference between the original and transformed dataset is obtained using the value difference (VD). To quantify the change in value positions, rank position (RP), rank maintenance (RK), change of rank of features (CP), and maintenance of rank of features (CK) are computed. Details on these PMs are defined in [16].

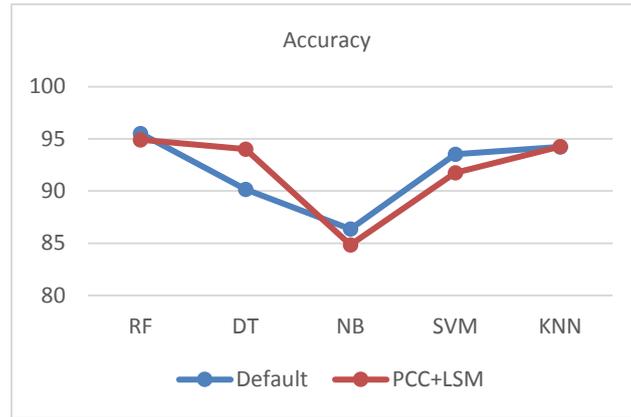

**Fig. 6.** Comparison of accuracy.

Fig. 6 presents the DU measure achieved before and after application of the PCC-LSM model for the RF, DT, NB, SVM and KNN classifiers. It can be observed that performance accuracy is similar with slight variations. DT and KNN achieved a higher accuracy with PCC-LSM model compared to the default set up. Examining the performance accuracy of the classifiers, it can be concluded that the distorted data retained the quality of the original data and hence, the PCC-LSM model provided good data quality.

**Table 2.** Privacy measures for proposed PCC-LSM.

|  | VD | RP | RK | CP | CK | Time (sec) |
|---|---|---|---|---|---|---|
| **LSM** | 1.11 | 30454.83 | 0.53 | 12.43 | 0.02 | 5.26 |
| **PCC + LSM** | **1.11** | **25565.82** | **0.61** | **8.07** | **0.0** | **3.54** |

Table 2 illustrates the PMs obtained on the UNSW-NB15 dataset. The value difference (VD) is almost similar for both LSM model and PCC-LSM model following the data distortion process, which implies that there is the same amount of data loss. Also, a high VD value indicates that there is no correlation between the original and distorted data and thus, offers more privacy to the network intrusion dataset. The large values of RP and CP along with small values for RK and CK infers that the original dataset is highly distorted and therefore, the privacy of the data is preserved. When comparing the privacy measures between the LSM model and proposed PCC-LSM model, it is observed that the RP and CP values for LSM is higher and RK value is lower. This is because PCC-LSM with eliminated features reduces the data complexity which results in smaller privacy metrics. Yet, by the definition of RP, RK, CP and CK the privacy of the data is considered strong. It is seen that the PCC-LSM model with fewer attributes has an improved distortion time of 3.54 seconds compared to 5.26 seconds of LSM. Reducing the number of features in PCC reduces the number of iterations in the distortion mechanism which reduces the time required for distortion.

## 5 Conclusion and Future Work

This paper investigates the fusion of the PCC and LSM techniques for low-cost intrusion detection while preserving user privacy. The PCC technique is initially applied to the UNSW-NB15 training dataset to reduce the dimensionality of the problem space, followed by the implementation of the LSM to transform the original dataset into a format that preserves privacy to prevent disclosure (accidental or otherwise) of sensitive/private data. The distorted UNSW-NB15 training dataset is then evaluated using the RF, DT, SVM, NB and KNN classifiers. RF, DT and KNN achieves similar or better results compared with SVM and NB in terms of Precision, Specificity, Accuracy and F-Score, except Recall. The performance of the proposed PCC-LSM-NIDS for intrusion detection reduces the overall computational time. Additionally, PCC-LSM-NIDS provides a reasonable level of user privacy, observed from the experimental results.

With good intrusion detection and privacy protection results achieved using the proposed PCC-LSM model, research may be continued to further reduce the computation time and increasing the accuracy of NIDS using other feature selection and feature extraction techniques such as Recursive Feature Elimination (RFE), and Linear Discriminant Analysis (LDA) with new datasets such as TUIDS, DDoS and SNMP_MIB to validate the consistency of the feature selection and the proposed PCC-LSM-NIDS. Future research using deep learning algorithms can be undertaken to further enhance the performance of the system.